# Two-particle interference of electron pairs on a molecular level

M. Waitz[1], D. Metz[1], J. Lower[1], C. Schober[1], M. Keiling[1], M. Pitzer[2], K. Mertens[3], M. Martins[3], J. Viefhaus[4], S. Klumpp[5], T. Weber[6], H. Schmidt-Böcking[1], L. Ph. H. Schmidt[1], F. Morales[7], S. Miyabe[8], T. N. Rescigno[9], C. W. McCurdy[9,10], F. Martín[11,12,13], J. B. Williams[14], M. S. Schöffler[1], T. Jahnke[1], and R. Dörner[1, *]

[1] Institut für Kernphysik, J. W. Goethe Universität, Max-von-Laue-Str.1, 60438 Frankfurt, Germany
[2] Universität Kassel, Heinr.-Plett-Straße 40, 34132 Kassel, Germany
[3] Institut für Experimentalphysik, Universität Hamburg, Luruper Chaussee 149, 22761 Hamburg, Germany
[4] FS-PE, Deutsches Elektronen-Synchrotron DESY, Notkestrasse 85, 22607 Hamburg, Germany
[5] FS-FL, Deutsches Elektronen-Synchrotron DESY, Notkestrasse 85, 22607 Hamburg, Germany
[6] Chemical Sciences Division, Lawrence Berkeley National Laboratory, Berkeley, California 94720, USA
[7] Max-Born-Institut, Max Born Strasse 2A, D-12489 Berlin, Germany
[8] Attosecond Science Research Team, RIKEN Center for Advanced Photonics, 2-1 Hirosawa, Wako-shi, Saitama 351-0198, Japan
[9] Ultrafast X-ray Science Laboratory, Chemical Sciences Division, Lawrence Berkeley National Laboratory, Berkeley, California 94720, USA
[10] Department of Chemistry, University of California, Davis, California 95616, USA
[11] Departamento de Química, Universidad Autonoma de Madrid, 28049 Madrid, Spain
[12] Instituto Madrileño de Estudios Avanzados en Nanociencia, 28049 Madrid, Spain
[13] Condensed Matter Physics Center (IFIMAC), Universidad Autonoma de Madrid, 28049 Madrid, Spain
[14] Department of Physics, University of Nevada Reno, 1664 N. Virginia Street Reno, NV 89557, USA

*We investigate the photo-doubleionization of $H_2$ molecules with 400 eV photons. We find that the emitted electrons do not show any sign of two-center interference fringes in their angular emission distributions if considered separately. In contrast, the quasi-particle consisting of both electrons (i.e. the "dielectron") does. The work highlights the fact that non-local effects are embedded everywhere in nature where many-particle processes are involved.*

**Introduction**: The two most counterintuitive cornerstones of quantum mechanics are the superposition principle giving rise to interference phenomena, and entanglement between distinct particles establishing what Einstein called "spooky action at a distance". A phenomenon which can only be explained by combining both these effects is two-particle interference [1], where the detection of an individual particle does not show interference fringes, but a coincidence measurement of two particles traversing a double slit array does. It has been demonstrated for entangled photon pairs in many experiments [2]. Here we report on the observation of this non-classical phenomenon for a pair of electrons emitted from a molecule through photo-double-ionization.

If a single particle can reach its final position along two indistinguishable pathways interference occurs, which may lead to the extinction of particle flux in directions where a single pathway would yield flux. This has been demonstrated in experiments on photons, neutrons [3], electrons [4], atoms, molecules [5, 6], and clusters [7]. Two-particle interference [1, 2] refers to the situation that both partners of an entangled pair are each sent separately into double slits. It can then occur that neither of the two particles individually shows any sign of interference, but that a coincidence measurement of both particles does. This highly non-classical effect was proposed in 1989 [1] and has been shown for photon pairs from parametric down conversion (see [2] for a review). For massive particles, the related Hong-Ou-Mandel effect has been recently demonstrated [8]. Here we reveal the existence of conditional two-particle interference between two electrons liberated through photoionization.

Figure 1A illustrates the principle of two-particle interference. Parent particles located within the source region decay into pairs of daughter particles. For parents at rest in the laboratory frame, the daughters are emitted back-to-back. Each daughter of an emission pair passes through one of two oppositely-opposed double-slit arrays before reaching detectors in far field. If the two double slits are coherently illuminated, single particle interference fringes are observed behind each pair of slits. To achieve coherent illumination the source region must be tightly localized. Under this condition the uncertainty principle ensures that the transverse spread in the momenta of the two daughters is sufficiently large that determination of the slit through which one passes does not establish the slit through which the other passes. Conversely, if the source extension is large then the single-particle interference patterns are lost. In this case determination of the slit through which one daughter passes establishes through which diametrically-opposed slit the other passes. However, under these same source conditions, entanglement between the birth positions of daughter particles emerging at a location $d$ along the $y$-axis allows strong two-particle interference fringes to be observed.

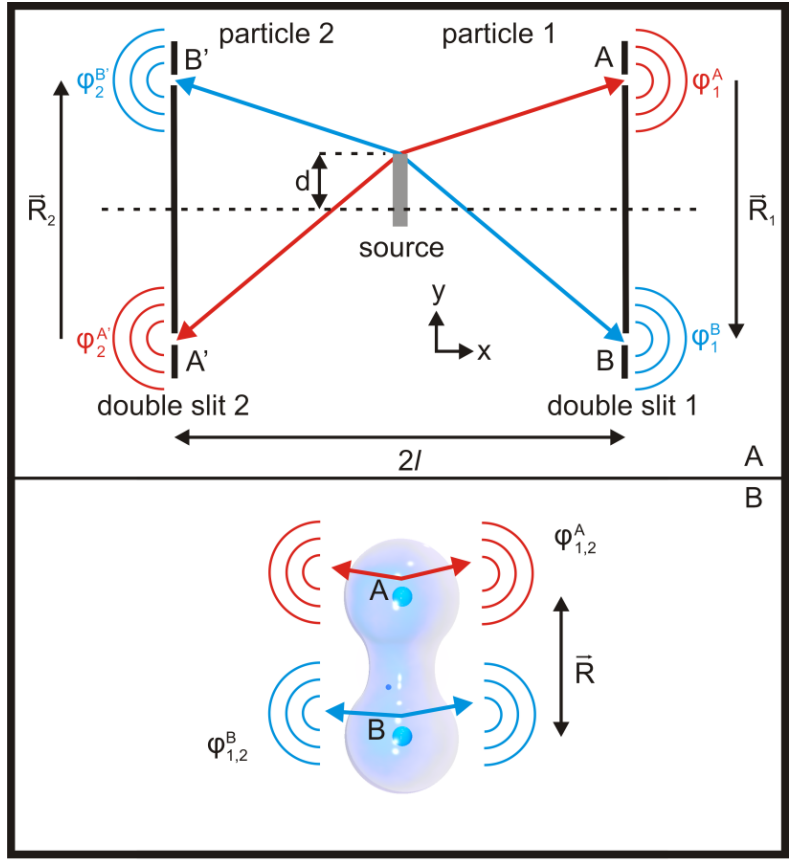

**FIG. 1.** A: Two-particle interference. An entangled pair of particles 1, 2 is emitted at a location $d$ along the $y$-axis within an extended source. Particle 1 travels to the double slit 1 on the right where two waves $\varphi_1^A$ and $\varphi_1^B$ emerge. The second particle travels to the left giving rise to the waves $\varphi_2^{A'}$ and $\varphi_2^{B'}$. Single particle detection in the far field of either double slit does not show any interference fringes if the source is sufficiently extended along the $y$-axis. Coincidence detection of the particle pair revives the interference (figure inspired by [9]). B: Implementation of the two-particle interference scheme for electrons. An electron pair, created by photo double ionization of $H_2$ and described by the two-electron waves $\varphi_{1,2}^A$ and $\varphi_{1,2}^B$, emerges from the two indistinguishable centers $A$ or $B$ separated by an internuclear distance $R$. The symbol $l$ represents the separation between source and double slits measured along the $x$-axis.

In our experiment we create an analogous situation for an *electron pair* emitted by the process of single-photon double ionization of $H_2$ at 400eV photon energy. The experiment has been performed at beamline P04 of the synchrotron PETRAIII in Hamburg, Germany using a COLTRIMS Reaction Microscope [10, 11]. Circular polarized photons of 400 eV photon energy are crossed with a supersonic molecular gas jet of $H_2$ in the center of a COLTRIMS spectrometer. Electrons and ions are guided to opposite sides by an electric field of 92 V/cm and a parallel magnetic field of 35.5 Gauss towards two position sensitive micro-channel plate detectors with a hexagonal delay line readout [12]. The spectrometer comprised an electron arm with an acceleration region of 3.7 cm and an ion arm with 5.5 cm acceleration region and 11.0 cm drift region. These settings yielded a $4\pi$ collection solid angle for electrons up to 420 eV and protons up to 28 eV. From the times-of-flight and positions of impact the momentum vectors of all particles are determined. The orientation of the inter-nuclear axis and the inter-nuclear distance at the instant of two-electron emission is obtained, event by event, from the momentum vectors of the two protons which fly apart almost back-to-back [13] with a kinetic energy release (KER) given by KER = 1/R (in atomic units). To avoid any bias dead time effects from our electron detector we have analyzed only those events where one electron was detected. The momentum vector of the second electron has been calculated from the momenta of the detected electron and two protons using momentum conservation.

At such high photon energies, much larger than the binding energy, the photo absorption occurs very close to either nucleus. Furthermore, we post-select only electron pairs for which the two electrons have similar energies. These pairs derive predominantly through a two-step process, where the absorption of the photon by one of the electrons is followed by a hard binary collision with the other one [14]. Under these conditions two-electron emission can be described by two-electron waves emerging separately from the locations of the two hydrogen nuclei. The idea that electron emission from a homonuclear diatomic molecule mimics double slit interference goes back to Cohen and Fano [15, 16]. This has been experimentally confirmed for single- and double-photoionization [17-19], for ionization by ion- and electron-impact [20] and for Auger electron emission [21, 22]. Those studies have shown that single particle interference is independent of the details of the ionization process and depends only on the two-center nature of the target. Two-particle interference, on the other hand, additionally requires entanglement of the electron pair and its preservation, both requirements being satisfied for the ionization process considered here. However, we note that for more complicated ionization processes (e.g. those involving sequential ionization in strong field ionization, charged particle impact or autoionization) this second requirement may not necessarily be fulfilled. Consequently, two-particle interference may be suppressed or even extinguished in those cases.

In Fig. 1B we show the two-electron emission process schematically. The two nuclei, labelled $A$ and $B$ act as sources of waves, as do the slits in a double slit arrangement. However, in contrast to the scenario of Fig. 1A, our two "double slits" are superimposed upon one another i.e. $A = A', B = B'$.

In Fig. 1A the waves emerging from the double slits are drawn as spherical waves. Electrons however exhibit strong mutual interaction, and in our scenario electrons emerge simultaneously through a common center. Thus the amplitudes $\varphi_{1,2}^A$ and $\varphi_{1,2}^B$ representing two-electron emission from sites $A$ and $B$, shown in Fig. 1B, cannot be expressed as products of two spherical waves, one describing each particle. Nonetheless, for far-field observation $|\vec{r}_{1i}|, |\vec{r}_{2i}| \gg R$, where $i = A, B$ and $\vec{r}_{1i}$ and $\vec{r}_{2i}$ represent electron position vectors originating from sites $A$ and $B$ respectively to the point of observation, $\varphi_{1,2}^A$ and $\varphi_{1,2}^B$ differ only by their different point of origin. The path differences experienced by waves $\varphi_{1,2}^B$ and $\varphi_{1,2}^A$ in the asymptotic region determines their relative phase at the points of detection. This far-field phase difference is simply $e^{i(\vec{k}_1+\vec{k}_2)\cdot\vec{R}}$. We can therefore express the probability amplitude describing the emission of an electron pair from the upper nucleus of Fig. 1B as $A(\vec{k}_1, \vec{k}_2)$, and that describing pair-emission from the lower nucleus as $A(\vec{k}_1, \vec{k}_2) \cdot e^{i(\vec{k}_1+\vec{k}_1)\cdot\vec{R}}$. These amplitudes describe emission through two indistinguishable paths; the two-electron emission probability $|\psi_{2e}|^2$ is therefore:

$$|\psi_{2e}|^2 \propto \left|A(\vec{k_1},\vec{k_2})\right|^2 \cdot \cos^2\left[\left(\frac{\vec{k_1}+\vec{k_2}}{2}\right)\cdot\vec{R}\right]. \tag{1}$$

Eqn. 1 predicts strong interference fringes in the two-electron emission probability emerging through the term $\cos^2\left[\left(\frac{\vec{k_1}+\vec{k_2}}{2}\right)\cdot\vec{R}\right]$. In contrast, $\left|A(\vec{k_1},\vec{k_2})\right|^2$ is a smoothly-varying function of $\vec{k_1}$ and $\vec{k_2}$. It incorporates the physics of single photon double ionization from a single center.

The key feature to the interference term $\cos^2\left[\left(\frac{\vec{k_1}+\vec{k_2}}{2}\right)\cdot\vec{R}\right]$ is the sum momentum $\vec{k_{sum}} = \vec{k_1} + \vec{k_2}$ of the two emitted electrons. Accordingly, the case of photo double ionization can be understood as the emission of a *dielectron* quasi-particle. This picture combining two emitted electrons (i.e. two particles that are, for example, located at different positions in space after the emission process) into a single quasi-particle accords strongly with our experimental results.

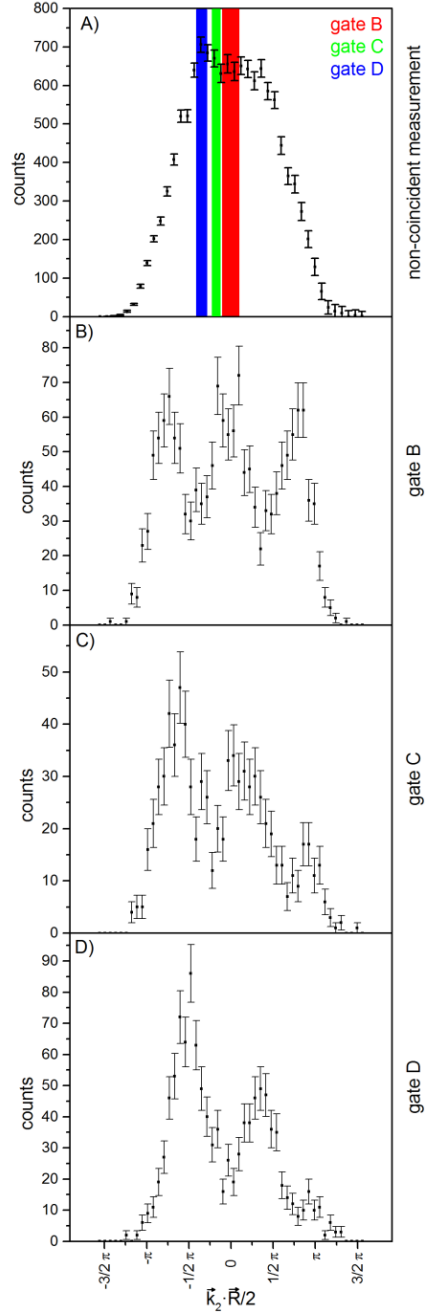

**FIG. 2.** Observation of conditional two-particle interference for an electron pair emitted by absorption of 400eV circular polarized photons at $H_2$ (Fig 1B). All panels correspond electron energies $E_1$ and $E_2$ where $0.15 < (E_1-E_2)/(E_1+E_2) < 0.85$. The horizontal axis shows $\vec{k_2} \cdot \vec{R}/2$ which is the component of electron momentum parallel to the molecular axis, scaled by half the internuclear distance. The data are integrated over all orientations of the photon propagation axis. A: electron 1 without selecting the second electron; the gates used to select events for panels B, C, D. B: coincident detection of electrons 1 and 2 with electron 1 post-selected in gate B as shown in panel A. C: same as B for selection of gate C. D: same as B for selection of gate D.

In our experiment we determine the two-electron momenta and the vector of the internuclear axis $\vec{R}$ for each photoionization event. The internuclear distance $R$ of the two atoms of the hydrogen molecule exhibits a finite spread determined by the vibrational ground state of the molecule. However, as $R$ is measured for each event in our experiment we choose to examine the scaled momentum $\vec{k} \cdot \vec{R}/2$ which compensates for the vibrational spread. We note that the data are integrated over all orientations of the polarization and light propagation making them insensitive to the choice of light polarization and minimizing the influence of the dipole character of the ionization process. In Fig. 2 we show accumulated counts as a function of $\vec{k_2} \cdot \vec{R}/2$, i.e. the scaled momentum of one of the two emitted electrons. The distribution depicted in panel 2A shows little sign of interference. Interference fringes emerge, however, when the momentum of the other electron is restricted to a certain value as shown in panels B, C, and D. In each of these panels we select the subset of coincidence events where electron 1 is detected in its respective gated region shown in panel A. The interference pattern is restored and occurs slightly shifted depending on the selected momentum value of electron 1. The origin of this effect can be clearly seen in Fig. 3. Here the scaled momenta of the two electrons are shown in a coincidence map. In this kind of plot, single particle interference fringes would show up as vertical and horizontal lines. Instead, a prominent diagonal feature is observed. A feature occurring along a diagonal with a slope of 45° belongs to the sum of the two quantities plotted along the x- and the y-axis. Accordingly, fringes along this direction correspond to transmission of the aforementioned fictitious dielectron quasi-particle of wave vector $\vec{k_1} + \vec{k_2}$ passing through a pair of slits. The rhombus-shaped envelope reflects the constraint imposed on the individual electron energies by energy conservation which is incorporated in the term $|A(\vec{k_1}, \vec{k_2})|^2$. With the sum momentum being at the heart of the phenomenon, the partial recovery of the interference fringes in $\vec{k_2} \cdot \vec{R}/2$ for different fixed values of $\vec{k_1} \cdot \vec{R}/2$ becomes obvious: as one summand is specified, the distribution of the other summand will show the overall features of the sum. By using $\vec{k_1} = 0$ one can furthermore explain the *one-particle* interference patterns observed in photo-double ionization of $H_2$ when one of the electrons takes most of the available energy and the slow electron is disregarded [17, 23]. A similar coincidence map (Fig. 3B) has been obtained from nearly exact theoretical calculations performed for a fixed internuclear distance of 1.4 a.u. and a photon energy of 375 eV. Very good agreement between theory and experiment is observed.

The calculations have been performed by using the method described in [24] and successfully used in [23] to evaluate double ionization cross sections at high photon energies. Briefly, we have used the exterior complex scaling (ECS) method implemented with the discrete variable representation (DVR) in finite elements for the radial variables of each of the two electrons. The radial grid extends up to $90a_0$ and the exterior scaling branching point was set at $50a_0$. Typically, the grid contained 209 DVR polynomial basis functions for each electron. We have used a one-center expansion of the two-electron wave function around the center of the molecule in terms of products of spherical harmonics and we have included all such products with angular momenta up to $l = 9$. Convergence was checked by varying the parameters of the grid and the number of angular momenta included in the one-center expansion. We have produced fully differential cross sections for all electron energy sharings in intervals $\Delta(E_1 - E_2)/(E_1 + E_2) = 0.05$ and all molecular orientations and electron ejection directions in angular intervals $\Delta\theta = 10°$ and $\Delta\phi = 10°$, where $\theta$ and $\phi$ are the polar and azimuthal angles, respectively. Then, at each point of a dense 2D grid

($\overrightarrow{k_{1,i}} \cdot \vec{R}, \overrightarrow{k_{2,j}} \cdot \vec{R}$), the fully differential cross sections were numerically integrated over all possible molecular orientations and electron energy sharings from 0.15 to 0.85 by using rectangle rule.

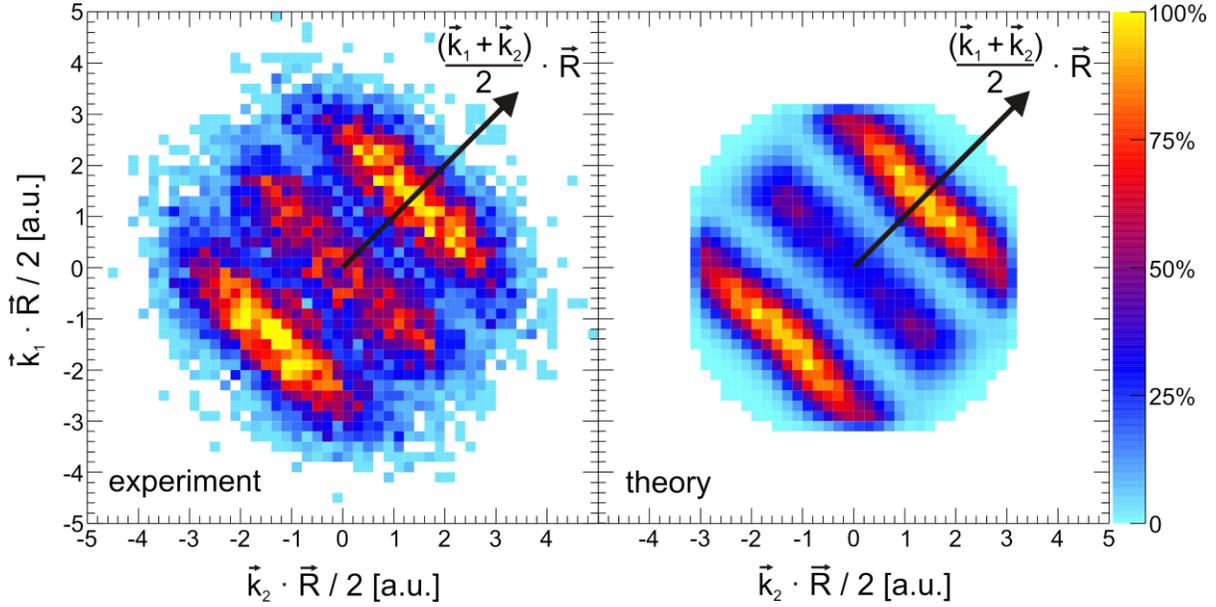

**FIG. 3. Correlations between electron momenta of the electron pair. Left panel, experiment. Right panel, ab initio theory. Horizontal and vertical axes show the scaled single-particle coordinates $\overrightarrow{k_1} \cdot \vec{R}/2$ and $\overrightarrow{k_2} \cdot \vec{R}/2$. The corresponding momentum component of the electron-pair center-of-mass is directed along the diagonal. Fig. 2A shows a projection of the data in this figure onto the horizontal axis. The experimental data is diagonally mirrored for better visual inspection. Highest intensity corresponds to 35 counts per bin.**

Inspection of the sum momentum in figure 3 provides new insight into the analogous optical two slit interference (Fig 1A). In this case the one double-slit-defining vector $\vec{R}$ in Fig 1B is split in two vectors $\vec{R}_1$ and $\vec{R}_2$ of equal magnitude and opposite sign due to the quasi back-to-back emission of the pair. By defining $\vec{R} = \vec{R}_1$ and using a similar approach to that employed in the derivation of Eqn. 1, it can be shown that the optical two slit interference for a particle pair emitted at a source location $d$ is determined by the scaled sum momenta $(\overrightarrow{k_1} + \overrightarrow{k_2})/2$ through the expression $cos^2\left[\left(\frac{\overrightarrow{k_1}+\overrightarrow{k_2}}{2}\right) \cdot \vec{R} + \phi\right]$. Here $\phi$ is a phase factor which depends on the location of emission $d$ and on the sum $k_1 + k_2$, where $k_i = |\vec{k}_i|; i = 1,2$. For the case $d = 0$ applying to our molecular double slit, $\phi = 0$. Thus one arrives at an expression identical in form to Eqn. 1, corresponding to the diffraction of a fictitious photon, of wavevector $(\overrightarrow{k_1} + \overrightarrow{k_2})/2$, by a single double slit.

We have demonstrated two-particle interference between interacting massive particles. It emerges naturally from photo fragmentation of molecules. The work highlights the fact that non-local effects are embedded everywhere in nature where many-particle processes are involved. Photoionization is just one way to project the effects of entanglement to the continuum where it can be detected.

**Acknowledgements:** This work was funded by the Deutsche Forschungsgemeinschaft, the BMBF, the European COST Action XLIC CM1204, the European Research Council Advanced Grant XCHEM no. 290853, and the MINECO Project no. FIS2013-42002-R. J.L. would like to thank DFG for support. We are grateful to the staff of PETRA III for excellent support during the beam time. Raw data are archived at the Goethe-University Frankfurt am Main and are available on request.

---

* doerner@atom.uni-frankfurt.de

[1] M. Horne, A. Shimony, and A. Zeilinger, *Phys. Rev. Lett*. **62**, 2209-2212 (1989).
[2] J.-W. Pan, Z.-B. Chen, C.Y. Lu, H. Weinfurter, A. Zeilinger, and M. Zukowski, *Rev. Mod. Phys.* **84**, 777-838 (2012).
[3] H. Rauch and S. A. Werner, *Neutron Interferometry*, Clarendon Press, Oxford (2000).
[4] C. Jönnson, *Zeitschrift für Physik* **161,** 454-474 (1961).
[5] M. Arndt, O. Nairz, J. Voss-Andreae, C. Keller, G. van der Zouw, and A. Zeilinger, *Nature* **401,** 680-682 (1999).
[6] T. Juffmann, H. Ulbricht, and M. Arndt, *Rep. Prog. Phys*. **76,** 086402 (2013).
[7] W. Schöllkopf and J.P. Toennies, *Science* **266**, 1345 (1994).
[8] R. Lopes, A. Imanaliev, A. Aspect, M. Cheneau, D. Boiron, and C.I. Westbrook, *Nature* **520,** 66 (2015).
[9] D.M. Greenberger, M.A. Horne, and A. Zeilinger, *Physics Today* **22,** 46 (1993).
[10] R. Dörner, V. Mergel, O. Jagutzki, L. Spielberger, J. Ullrich, R. Moshammer, and H. Schmidt-Böcking, *Physics Reports* **330,** 96-192 (2000).
[11] J. Ullrich, R. Moshammer, A. Dorn, R. Dörner, L. Ph. Schmidt, and H. Schmidt-Böcking, *Rep. Prog. Phys.* **66,** 1463-1545 (2003).
[12] O. Jagutzki, A. Cerezo, A. Czasch, R. Dörner, M. Hattass, M. Huang, V. Mergel, U. Spillmann, K. Ullmann-Pfleger, Th. Weber, H. Schmidt-Böcking, and G.D.W. Smith, *IEEE Transact. On Nucl. Science* **49,** 2477 (2002).
[13] T. Weber, A.O. Czasch, O. Jagutzki, A. K. Müller, V. Mergel, A. Kheifets, E. Rotenberg, G. Meigs, M.H. Prior, S.Daveau, A. Landers, C.L. Cocke, R. DiezMuino T. Osipov, H. Schmidt-Böcking, and R. Dörner, *Nature* **431,** 437-440 (2004).
[14] A. Knapp, A. Kheifets, I. Bray, Th. Weber, A.L. Landers, S. Schössler, T. Jahnke, J. Nickles, S. Kammer, O. Jagutzki, L.Ph. Schmidt, T. Osipov, J. Rösch, M.H. Prior, H. Schmidt-Böcking, C.L. Cocke, and R. Dörner, *Phys. Rev. Lett*. **89,** 033004 1-4 (2002).
[15] H.D. Cohen and U. Fano, *Phys. Rev*. **150,** 30-33 (1966).
[16] J. Fernandez, O. Fojon, A. Palacios, and F. Martin, *Phys. Rev. Lett*. **98,** 043005-1 -043005-4 (2007).
[17] D. Akoury, K. Kreidi, T. Jahnke, Th. Weber, A. Staudte, M. Schöffler, N. Neumann, J. Titze, L. Ph. H. Schmidt, A. Czasch, O. Jagutzki, R. A. Costa Fraga, R. E. Grisenti, R. Diez Muino, N. A. Cherepkov, S. K. Semenov, P. Ranitovic, C. L. Cocke, T. Osipov, H. Adaniya, J. C. Thompson, M. H. Prior, A. Belkacem, A. Landers, H. Schmidt-Böcking, and R. Dörner, *Science* **319,** 949 (2007).
[18] R.K. Kushawaha, M. Patanen, R. Guillemin, L. Journel, C. Miron, M. Simon, M.N. Piancastelli, C. Skates, and P. Decleva, *Proc Natl Acad Sci USA*. **110,** 15201 (2013).

[19] S.E. Canton, E. Plesiat, J.D. Bozek, B.S. Rude, P. Decleva, and F. Martin, *Proc Natl Acad Sci USA* **108,** 7302 (2011).
[20] N. Stolterfoht, B. Sulik, V. Hoffmann, B. Skogvall, J.Y. Chesnel, J. Rangama, F. Fremont, D. Hennecart, A. Cassimi, X. Husson, A.L. Landers, J.A. Tanis, M.E. Galassi, and R.D. Rivarola, *Phys. Rev. Lett*. **87,** 023201 (2001).
[21] X.-J. Liu, Q. Miao, F. Gel'mukhanov, M. Patanen, O. Travnikova, C. Nicolas, H. Agren, K. Ueda, and C. Miron, *Nat. Photonics* **9,** 120 (2015).
[22] N. A. Cherepkov, S. K. Semenov, M. S. Schöffler, J. Titze, N. Petridis, T. Jahnke, K. Cole, L. Ph. H. Schmidt, A. Czasch, D. Akoury, O. Jagutzki, J. B. Williams, T. Osipov, S. Lee, M. H. Prior, A. Belkacem, A. L. Landers, H. Schmidt-Böcking, R. Dörner, N. A. Cherepkov, S. K. Semenov, and Th. Weber, *Phys. Rev*. **82,** 023420 (2010).
[23] D. A. Horner, S. Miyabe, T. N. Rescigno, C. W. McCurdy, F. Morales, and F. Martín, *Phys. Rev. Lett.* **101**, 183002 (2008).
[24] W. Vanroose, D. A. Horner, F. Martín, T. N. Rescigno, and C. W. McCurdy, *Phys. Rev. A* **74**, 052702 (2006).